\documentclass[conference,10pt]{IEEEtran}

 \IEEEoverridecommandlockouts
\usepackage{amssymb}
\usepackage{mathrsfs}
\usepackage{amsfonts}
\usepackage{amsmath}
\usepackage{footnote}
\usepackage{txfonts}
\usepackage{algorithm}
\usepackage{algorithmic}
\usepackage{tabularx}

\usepackage{cite}

\usepackage[dvips]{graphicx}

   \DeclareGraphicsExtensions{.eps,.pdf}

\ifCLASSOPTIONcompsoc
\usepackage[tight,normalsize,sf,SF]{subfigure}
\else
\usepackage[tight,footnotesize]{subfigure}
\fi

\begin{document}
\title{Joint Relay Selection and Link Adaptation for Distributed Beamforming in Regenerative Cooperative Networks
\thanks{This paper is supported by the National Science
Foundation of China (NSFC 60702051, NSFC-AF: 60910160), the
Specialized Research Fund for the Doctoral Program of Higher
Education (SRFDP£º20070013028), and the Program for New Century
Excellent Talents in University (NCET-08-0735). This paper is
co-funded by Nokia on the Beyond 3G Research project. }}

\author{\IEEEauthorblockN{Wei Yang\IEEEauthorrefmark{1}, Lihua Li\IEEEauthorrefmark{1}, Gang Wu\IEEEauthorrefmark{2}, Haifeng Wang\IEEEauthorrefmark{2}}
\IEEEauthorblockA{\IEEEauthorrefmark{1}Key Lab. of Universal Wireless Commun., Beijing University of Posts and Telecom.(BUPT)\\
Ministry of Education, Wireless Technology Innovation
Institute(WTI), BUPT, China}
\IEEEauthorblockA{\IEEEauthorrefmark{2}Wireless Modem System
Research, Device R\&D, NOKIA, Shanghai, China} }

\maketitle

\begin{abstract}
Relay selection enhances the performance of the cooperative networks
by selecting the links with higher capacity. Meanwhile link
adaptation improves the spectral efficiency of wireless data-centric
networks through adapting the modulation and coding schemes (MCS) to
the current link condition. In this paper, relay selection is
combined with link adaptation for distributed beamforming in a
two-hop regenerative cooperative system. A novel signaling mechanism
and related optimal algorithms are proposed for joint relay
selection and link adaptation. In the proposed scheme, there is no
need to feedback the relay selection results to each relay. Instead,
by broadcasting the link adaptation results from the destination,
each relay will automatically understand whether it is selected or
not. The lower and upper bounds of the throughput of the proposed
scheme are derived. The analysis and simulation results indicate
that the proposed scheme provides synergistic gains compared to the
pure relay selection and link adaptation schemes.

\end{abstract}

\begin{IEEEkeywords}
relay selection, link adaptation, distributed beamforming,
signaling, throughput.
\end{IEEEkeywords}

%
\IEEEpeerreviewmaketitle

\section{Introduction}
Cooperative communication networks, in which wireless nodes
cooperate with each other in transmitting information, promise
significant gains in overall throughput and create robustness
against channel fading \cite{Laneman}\cite{patent}. A variety of
cooperative schemes have been proposed in the literatures with
different design issues and channel information assumptions.
\par
Link adaptation can help mitigate the effects of time-varying fading
channel as well as exploit favorable channel conditions when they
exist \cite{GoldSmith_AMC}. Link adaptation for cooperative networks
are considered in \cite{Zinan_GC} to maximize the data throughput.
\par
Independent of link adaptation, the network performance can also be
improved by selecting the cooperating relays to exploit temporal and
spatial diversity \cite{single_multiple_relay_selection}.
Relay selection simplifies signaling, avoids complex synchronization
schemes, and with careful design can preserve the spatial diversity
provided by the total number of relays in the network
\cite{A_Bletas}.
\par
For the networks with parallel relays, distributed beamforming (also
called network beamforming in \cite{Jing_AF}) is proved optimal if
perfect channel state information (CSI) or high quality channel
information feedback from the receiver is available at the
transmitter \cite{distributed_beamforming_Comm_Mag}. In this
approach, the relays linearly weight their transmit signals
according to the CSI so that they can add up coherently at the
destination. The beamforming problem for networks with limited
feedback from destination to relays is studied in \cite{Jing_YD_bit}
for non-regenerative relays, and in \cite{dis_BF_DF} for
regenerative relays.

\par
The combination of relay selection and link adaptation is proposed
in \cite{Jonit_Rate}. However, only single relay is selected, and
distributed beamforming is not considered in this scheme. Moreover,
it is designed for multi-hop transmission, in which relay and
modulation and coding scheme (MCS) are selected in each hop by the
forwarding relays (acting as the source in each hop).

\par
In this paper, joint relay selection and link adaptation are
considered for two-hop regenerative cooperative systems. The
modulation scheme and relays are selected for distributed
beamforming to forward data from the source to the destination.
Unlike that in \cite{Jonit_Rate}, the selection in our proposal is
performed by the destination, and the selection results are fed back
to the source and relays. To reduce the signaling overhead of the
relay selection and link adaptation, a novel signaling mechanism is
then proposed to broadcast the link adaptation results from the
destination, and each relay will automatically understand whether it
is selected or not. Moreover, a simple relay and modulation scheme
selection algorithm is designed with the complexity linear to the
number of candidate modulation schemes. The analysis and simulation
results of the throughput and symbol error rate (SER) show that the
proposed scheme outperforms pure relay selection and link adaptation
schemes.

\par
This paper is organized as follows. Section \ref{system_model}
presents the system model. Section \ref{Scheme} introduces the joint
relay selection and link adaptation scheme, and the optimal
selection algorithm. The lower and upper bounds of the throughput of
the proposed scheme are derived in Section \ref{Performance
Analysis}, and simulation results of the throughput and SER
performance are provided. Finally, conclusions are drawn in Section
\ref{conclusion}.

\section{System Model}
\label{system_model} 
\begin{figure}
\centering
\includegraphics[width=2.1in]{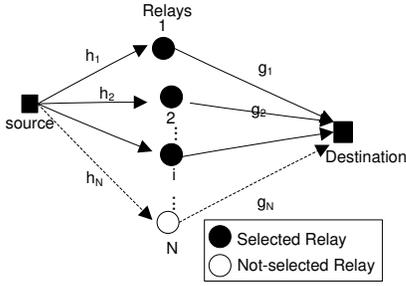}
\caption{ Wireless network with parallel relays.} \label{figure1}
\end{figure}

Consider a two-hop regenerative cooperative
system with a source node, a destination node and a set of $N$
candidate relays $\mathcal {R} = \{1,2,\cdots N\}$, as illustrated
in Fig \ref{figure1}. Each relay node has only one antenna which
cannot transmit and receive simultaneously. Assume there is no
direct link between the source and the destination. Denote the
channel from the source to the $i$-th relay as $h_i$ and the channel
from the $i$-th relay to the destination as $g_i$. $h_i$ and $g_i$
are assumed to be independent and modeled as
$\mathcal{C}\mathcal{N}(0,1)$. The received noise at each receiver
is normally distributed $\sim$ $\mathcal {C}\mathcal {N}(0,N_0)$.
The source transmits under an average power constraint $P_s$, and
relays are under a total power constraint, which is equal to $P_s$.
Denote $\rho$ as the signal-to-noise ratio (SNR) without fading,
i.e., $\rho=\frac{P_s}{N_0}$. Depending on the channel states, only
a subset, $\mathcal {A}\subset \mathcal {R}$ is selected to help the
transmission between the source and the destination. $A$ is used to
denote the cardinality of $\mathcal {A}$.
\par

In the first hop, the source node transmits $s$ to the relays, where
$s\in\mathcal {C}$, and $\mathcal {C}$ is a finite constellation
with average unit energy and $M$ cardinality. For simplicity of the
performance analysis, square $M$-ary quadrature amplitude modulation
($M$-QAM) is used as candidate \textbf{modulation scheme (MS)}.
However, the proposed scheme do not limit to this assumption, and
continue to work in a coded system with various MCS. Denote QAM
scheme with $2^{2k}$ cardinality as $MS_{(k)}$, $k=1,\cdots L$, and
$\mathcal{S}=\{MS_{(1)},MS_{(2)},...MS_{(L)}\}$ as the set of MSs.
The $i$th relay receives
\begin{equation}
r_i=\sqrt{P_s}h_i s+n_i,
\end{equation}
where $P_s$ is the average power used at the source, and $n_i$ is
the noise at relay node $i$. The $i$-th relay demodulates the
received signal as $\hat{s}_i$ using maximum likelihood (ML)
demodulation,
\begin{equation}
\hat{s}_i=\arg\min\limits_{s_i\in\mathcal{C}}\|r_i-\sqrt{P_s}h_i s\|.
\end{equation}
\par
In the second hop, the selected relays transmit simultaneously to
the destination using distributed beamforming. The beamforming
weight for the $i$-th selected relay is $\omega_i$,
$i\in\mathcal{A}$.
Assume the MS used in the second hop is the same as the first hop.
\footnote{Fixing modulation scheme simplifies the design of relays
and avoids complex signaling during the data transmission, although
using a different modulation schemes for the second phase
transmission can bring extra throughput gains \cite{patent}. } The
received signal at the destination is
\begin{equation}
y_d = \sum \limits_{i \in \mathcal{A}} \sqrt{P_s} g_i \omega_i \hat{s}_i + n_d,
\end{equation}
where $n_d$ is the noise at the destination.
\par
Given the knowledge of CSI, if all the selected relays demodulate
the signal from the source correctly, the optimal beamforming weight
at each selected relay $i$ is $\frac{g_ {i} ^ {\ast} } {\sum
\nolimits_ {i \in \mathcal{A}} {|g_i|^2}}$ \cite{patent}, where
$g_{i}^{\ast}$ is the conjugate value of $g_{i}$.

\section{Joint Relay Selection and Link Adaptation for Distributed Beamforming}
\label{Scheme}
\subsection{The general scheme}
\begin{figure}
\centering
\includegraphics[width=3in]{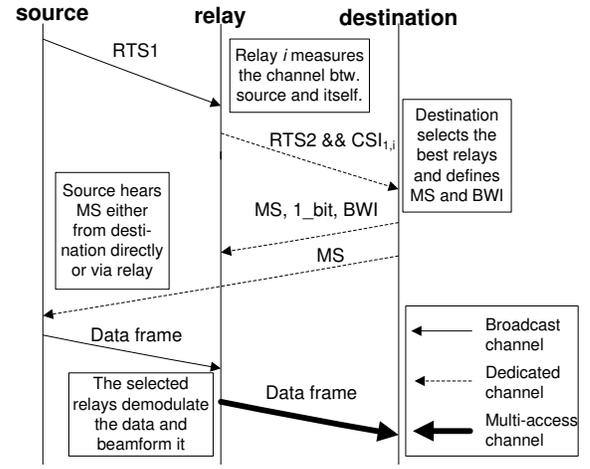}
\caption{ Flowchart of the proposed general joint relay selection
and link adaptation scheme} \label{figure2}
\end{figure}

This section presents a scheme for joint relay selection and link adaptation in a regenerative cooperative network, the flowchart of which
is shown in Fig. \ref{figure2}.
\par
The details of the proposed scheme are described as follows.
\par
\textbf{Step 1:} The source broadcasts a ready-to-send (RTS) packet,
denoted as RTS1. Each relay measures the CSI between source and
itself upon reception of RTS1, denoted as $CSI_{1,i}$.
\par
\textbf{Step 2:} Relays transmit the RTS2 packets containing
$CSI_{1,i}$ to the destination in their dedicated channels.
\par
\textbf{Step 3:} The destination selects the best relays based on
$|h_i|^2$ and $|g_i|^2$. It also decides the MS for the
transmission. Here the \emph{Best Throughput Criterion} is used
\begin{equation}
\label{criteria} [MS^{\ast},\mathcal {A}^{\ast}]=
\arg\max\limits_{MS, \mathcal {A}}\{\mathscr{R}(MS)|P_e(MS,\gamma_d)
\leq SER_{tgt}\},
\end{equation}
where $\mathscr{R}(MS)$ is the data rate of the corresponding
modulation scheme, $\gamma_d$ is the instantaneous received SNR at
the destination, $P_e(MS,\gamma_d)$ is the SER when the
corresponding MS is used, and $SER_{tgt}$ is the target SER. In the
following text, we denote $MS_i>MS_j$, if
 $\mathscr{R}\{MS_i\}>\mathscr{R}\{MS_j\}$.
The Best Throughput Criterion allows us to select the optimal MS and
relays, such that the throughput of the system is maximized while
the SER is below a certain target.

\textbf{Step 4:} The destination feeds back the selected
\textbf{MS}, the \textbf{relay selection result} (1 bit to indicate
each specific relay whether it is selected or not, denoted as 1\_bit
in Fig.~\ref{figure2}) and the \textbf{beamforming weight
information (BWI)} to each relay in its dedicated channel. The
source should be also notified the MS either from the destination
directly or via the selected relays. BWI is the weight information
to perform distributed beamforming at each relay. Different BWI
feedback mechanisms should be adopted in TDD and FDD systems.
\par
In TDD systems, the destination sends training sequence to relays.
The selected relays can estimate the channel gains $|g_i|$ based on
channel reciprocity. The destination only needs to broadcast the
summation of the channel power gains of the selected relays,
$\sum\nolimits_{i \in \mathcal{A}}|g_{i}|^2$ to the selected relays.
Each relay can adjust its transmission power into $\frac{|g_
{[i]}|^2 P_s } {(\sum \nolimits_{i \in \mathcal{A}}|g_{[i]}|^2)^2}$
accordingly. In FDD system, the destination has to feedback BWI
$\frac{g_ {i} ^ {\ast} } {\sum \nolimits_{i \in
\mathcal{A}}|g_{[i]}|^2}$ to each selected relay.

\par

\textbf{Step 5:} The source broadcasts data using the selected MS.
\par
\textbf{Step 6:} The selected relays demodulate the received signal
from the source and beamform it to the destination.
\par


\subsection{Improved feedback signaling}
Both relay selection and link adaptation require some information
exchange among source, relays and destination, which is a heavy
burden on signaling. Such as in IEEE 802.16j mobile relay (MR)
systems, the feedbacks from destination to relays are in each
relay's dedicated channel independently, and the feedback channel is
pre-scheduled by the scheduler in higher layer before relay
selection and link adaptation. Prior to scheduling, none of the
nodes know what relays will be actually selected. So the dedicated
signaling channel must be scheduled for all the relays, making the
overhead proportional to the number of relays in the network.
\par
\begin{figure}
\centering
\includegraphics[width=2.7in]{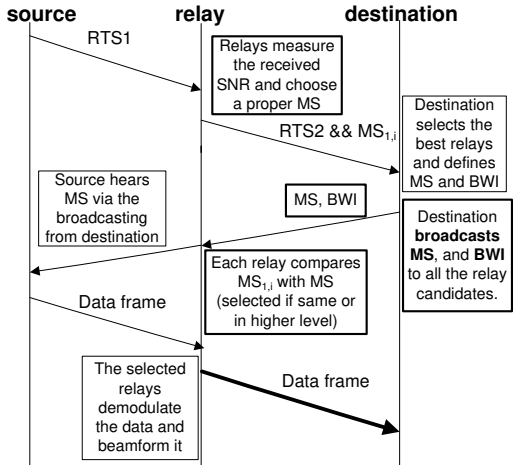}
\caption{ Flowchart of the proposed broadcasting signaling scheme
(vs. the general signaling scheme feedbacks MS, BWI and Relay Selection Index via dedicated channel).}
\label{figure3}
\end{figure}

To reduce the overhead, an improved feedback signaling for the joint
relay selection and link adaptation scheme is proposed. The
flowchart of the proposed feedback signaling is shown in Fig
\ref{figure3}.
 The process is described in details as follows:
 \par
\textbf{Step 1:} The source broadcasts a RTS1 packet. Upon reception
of RTS1, all the relays measure the SNR of the channel between
source and itself and define MS as $MS_{1,i}$.
 The MS selection criterion used in the proposed scheme is
\begin{equation}
MS_{1,i} =
\arg\max\limits_{MS}\{\mathscr{R}(MS)|P_e(MS,\gamma_{1,i})\leq
SER_{tgt}\},
\end{equation}
where $\gamma_{1,i}=\rho |h_i|^2$ is the SNR of the first hop at
each relay. In practical systems, MS can be determined by a
threshold based method, and the threshold between $MS_{(k)}$ and
$MS_{(k-1)}$ is
\begin{equation}
\Gamma (MS_{(k)}) = \arg\min\limits_{\gamma}\{\gamma |
Pe(MS_{k},\gamma)\leq SER_{tgt}\},k=1,\cdots L.
\end{equation}
\par
\textbf{Step 2:} Each relay sends its RTS2 packet with $MS_{1,i}$ to
the destination.
\par
\textbf{Step 3:} The destination then selects the relays, and
determines the MS for the first and second hop and BWI. The
selection criterion is the same as (\ref{criteria}).

\newcounter{MYtempeqncnt}
\begin{figure*}[!t]
\normalsize
\setcounter{MYtempeqncnt}{\value{equation}}
\setcounter{equation}{9}
\begin{equation}
\label{long}
f(u,v) = \begin{cases}
 \frac{{N!  }}{{\rho ^2 (N - A - 1)!(A - 1)!}} e^{ - \frac{(u+v)}{\rho }}\left( {1 - e^{ - \frac{v}{\rho }} } \right)^{N - A - 1}
 \cdot \left(
 e^{ - \frac{{Au}}{\rho }} \sum\limits_{j = 0}^{A - 1} {\frac{u^j  }{\rho ^j j!}}
  + \frac{u^{A - 1}}{\rho ^{A - 1} A!}
 \left( { e^{ - \frac{{Au}}{\rho }}+ e^{ - \frac{{Av}}{\rho }}-1} \right) \right),&0 < v < u \\
 \frac{{N!u^{A - 1}}}{{\rho ^{A + 1} \left( {N - A - 1} \right)!\left( {A - 1} \right)!A!}} e^{ - \frac{u}{\rho }}e^{ - \frac{{(A + 1)v}}{\rho }} \left( {1 - e^{ - \frac{v}{\rho }} } \right)^{N - A - 1} , &0 < u < v \\
 \end{cases}
\end{equation}

\setcounter{equation}{\value{MYtempeqncnt}}
\hrulefill
\vspace*{2pt}
\end{figure*}

\par
\textbf{Step 4:} The destination \textbf{broadcasts} the following
information to the source and relays: \textbf{MS} and \textbf{BWI}.
Note that, the destination does not need to feedback to each relay
the relay selection results. Instead, upon reception of MS from
destination, each relay compares it with $MS_{1,i}$, which is
determined in \textbf{Step 2}. The relay is selected if
\begin{equation}
MS_{1,i}\geq MS^{\ast},
\end{equation}
where $MS^{\ast}$ is the selected MS. Otherwise, the corresponding
relay candidates back off. The source can be notified about MS by
listening to the broadcasting from destination.
\par
\textbf{Step 5:} The source broadcasts data.
\par
\textbf{Step 6:}The selected relays demodulate the signal from the
source and beamform it to the destination.
\par

Note again in \textbf{Step 4}, the relay can know it is selected if
$MS^{\ast}$ is aligned with the link between it and source so that
there is no need for the destination to feedback the relay selection
result as in the general scheme. Overall, it can be noticed that the
signaling overhead can be considerably reduced by broadcasting
instead of dedicated signaling.

\subsection{The relays and MS selection algorithm}
To find the set $\mathcal{A}^{\ast}$ and modulation scheme
$MS^{\ast}$ in (\ref{criteria}), an exhaustive search would involve
over $L\cdot2^N$ cases. The computational complexity of the
exhaustive search grows exponentially with the number of relay
candidates. Below a simple algorithm to find the optimal set
$\mathcal{A}^{\ast}$ and modulation scheme $MS^{\ast}$ is proposed
as Algorithm 1.
\par
\begin{algorithm}
\caption{MS and Relay Selection Algorithm}
\begin{algorithmic}[1]
\STATE Sort $MS_{1,i}$ in descending order such that $MS_{1,\tau_1}\geq MS_{1,\tau_{2}}\geq \cdots \geq MS_{1,\tau_N}$, where $\tau_k \in \mathcal{R}$, $1\leq k \leq N$\\
\STATE $\gamma \leftarrow 0$, $i \leftarrow 1$ , $MS^{\ast} \leftarrow \mathrm{NULL}$, \\
 \WHILE{ $i \in \mathcal{R}$ \textbf{and}  $MS^{\ast} == \mathrm{NULL}$ }
 \IF{$MS_{1,\tau_i}\neq MS_{1,\tau_i+1}$}
 \STATE $\gamma=\rho \sum \nolimits_{j=1}^{i}|g_{\tau_j}|^2$\\
 \IF{$\gamma \geq \Gamma(MS_{1,\tau_i})$}
 \STATE $MS^{\ast} \leftarrow MS_{1,\tau_i}$\\
 \STATE $\mathcal{A}^{\ast} \leftarrow \{1,\cdots i\}$\\
 \ENDIF
 \ENDIF
 \STATE $i \leftarrow i+1$
\ENDWHILE
\end{algorithmic}
\end{algorithm}
In this algorithm, the selected relays are those which have the
largest source-relay channel SNR, among the set of all candidate
relays $\mathcal{R}$. The SNR at the destination when relays in
$\{1,\cdots \tau_i\}$ are selected is shown in the line 5 of the
Algorithm 1, assuming no demodulating error at the selected relays.
The SNR loss at the destination due to demodulating errors at the
selected relays is analyzed in Section \ref{Performance Analysis}.
It is clear that adding more relays to the set $\mathcal{A}$
increases the SNR at the destination, and hence is always beneficial
in the second hop transmission. That explains why the result of MS
selection can signal the results of relay selection.
\par
The computational complexity of the proposed algorithm is decreased
dramatically. Only less than $L$ cases are involved in addition to a
sort operation of $N$ integers.
\par



\section{Performance Evaluation}
\label{Performance Analysis}
\subsection{Performance Analysis}
The instantaneous received SNR at the destination in the second
phase can be expressed as,
\begin{equation}
\label{gamma_d1}
\begin{split}
\gamma_d &=\frac {\rho | \sum\nolimits_{i\in \mathcal{A}}\omega_i
g_i|^2 }
{1+\rho\sum\nolimits_{i\in \mathcal{A}}\omega_i g_i \mathbb{E} \left(|\delta_i|^2\right)}\\
&= \frac{\rho \sum\nolimits_{i\in \mathcal{A}}{|g_i|^2}} {1+ \rho
\mathord{\left/
 {\vphantom {\rho  {\sum\nolimits_{i \in \mathcal{A}} {|g_i |} ^2 }}} \right.
 \kern-\nulldelimiterspace} {\sum\nolimits_{i \in \mathcal{A}} {|g_i |} ^2 } \cdot
\left( \sum\nolimits_{i\in \mathcal{A}}{|g_i|^2 \mathbb{E}
\left(|\delta_i|^2\right)} \right)},
\end{split}
\end{equation}
where $\delta_i=\hat{s}_i-s$ is the demodulating error of the $i$th
relay. The expectation of the demodulating errors at the $i$th relay
using square MQAM can be approximated as,
\begin{equation}
\label{E_delta}
\begin{split}
\mathbb{E}\left( |\delta _i |^2\right) & \approx \frac{{4(e_{\min } )^2 (\sqrt M  - 1)}}{{\sqrt M }}Q\left( {\sqrt {\frac{{P_s \left(e_{\min }\right)^2 |h_i |^2}}{{2N_0 }}} } \right)\\
&= \frac{24}{M+\sqrt{M}} \cdot Q\left( {\sqrt
{\frac{3\gamma_{1,i}}{(M-1)} }}\right),
\end{split}
\end{equation}
where $e_{\min }=\sqrt{\frac{6}{M-1}}$ is the minimum Euclidean
distance of MQAM with unit average energy. Substituting
(\ref{E_delta}) into (\ref{gamma_d1}), the lower bound of the
received SNR can be obtained as
\begin{equation}
\label{eq10} \gamma_d \geq\sum\limits_{i \in
\mathcal{A}}{\gamma_{2,i}}- \underbrace{\sum\limits_{i \in
\mathcal{A}}{\frac{24 \gamma_{2,i}}{M+\sqrt{M}} \cdot Q\left( \sqrt
{\frac{3\gamma_{1,i}}{(M-1)} }\right)}}_{\text{SNR loss}},
\end{equation}
where $\gamma_{2,i}=\rho |g_i|^2$. The second term in (\ref{eq10})
is the SNR loss due to the demodulating errors at the selected
relays.
\par
The average throughput for the joint link adaptation and relay
selection schemes can be defined as
\begin{equation}
\label{throughput_eq}
\begin{split}
\zeta &= \frac{1}{2}\mathbb{E}\left[\mathscr{R}\left(MS\right)\left(1-BLER\right)\right]\\
&=\sum\limits_{k=1}^{L}{\sum\limits_{\mathcal{A}\subset\mathcal{R}}
{k \cdot
\mathrm{Pr}\left(\mathcal{A},MS_{(k)}\right)\left(1-P_e\left(\mathcal{A},MS_{(k)}\right)\right)^{L_{bit}/2k}}},
\end{split}
\end{equation}
where $BLER$ is the block error rate,
$\mathrm{Pr}\left(\mathcal{A},MS_{(k)}\right)$ is the probability
that the set $\mathcal{A}$ of relays and modulation scheme
$MS_{(k)}$ are selected, $P_e\left(\mathcal{A},MS_{(k)}\right)$ is
the SER and $L_{bit}$ is the length of each transmitted block.
\par
Using symmetry arguments, it can be concluded that
$\mathrm{Pr}\left(\mathcal{A},MS_{(k)}\right)$ and
$P_e\left(\mathcal{A},MS_{(k)}\right)$ are the same for all sets
$\mathcal{A}$ with the same cardinality. Hence, we have
$P_e\left(A,MS_{(k)}\right)=P_e\left(\mathcal{A},MS_{(k)}\right)$
and $\mathrm{Pr}\left(A,MS_{(k)}\right) = \dbinom{N}{A}\cdot
\mathrm{Pr}\left(\mathcal{A},MS_{(k)}\right)$.
\par
Sort the SNR of the source-relay channel in descending order $\gamma_{1,[1]}>\cdots>\gamma_{1,[N]}$. For simplicity of expression, denote
$\gamma_{[i]}=\gamma_{1,[i]}$. The probability that exact $A$ relays ($A<N$) and the $k$-th MS are selected is given by
\begin{equation}
\label{eq9}
\begin{split}
&\mathrm{Pr}\left(A,MS_{(k)}\right)\\
&=\mathrm{Pr}\left( \left\{ \Gamma_{(k)}\leq \min\left\{
\sum\limits_{i=1}^{A}{\gamma_{2,i}}, \gamma_{[A]}\right\} <
\Gamma_{(k+1)} \right\}\bigcap
\biggl\{\gamma_{[A+1]}<\Gamma_{(k)}\biggr\} \right),
\end{split}
\end{equation}
where $\Gamma_{(k)}= \Gamma(MS_{(k)})$, $k=1,\cdots,L$, and
$\Gamma_{(L+1)}= + \infty$. Define $U_A=\min\left\{
\sum\limits_{i=1}^{A}{\gamma_{2,i}}, \gamma_{[A]}\right\}$ and
$V_A=\gamma_{[A+1]}$. The joint probability density function (pdf)
of $U_A$ and $V_A$ can be derived as in (\ref{long}) at the top of
the page.
The proof of this result is given in Appendix \ref{appendixA}.
\par
Using the closed-form expression for $f(u,v)$ in (\ref{long}), the
expression for $\mathrm{Pr}\left(A,MS_{(k)}\right)$ can be written
as follows: \addtocounter{equation}{1}
\begin{equation}
\begin{split}
\label{pr_a}
&\mathrm{Pr}\left(A,MS_{(k)}\right)
\\& = \int_{0}^{\Gamma_{(k)}} \int_{\Gamma_{(k)}}^{\Gamma_{(k+1)}}{f(u,v)dudv}\\
&=\frac{N!}{A!} \sum\limits_{i=0}^{N-A}{\frac{(-1)^i
e^{-\tilde{\Gamma}_{(k)}i}}{i!(N-A-1)!}}\cdot
\left(e^{-(A+1)\tilde{\Gamma}_{(k)}}
\sum\limits_{j=0}^{A-1}{\frac{\left(\tilde{\Gamma}_{(k)}\right)^j}{j!}}\right.\\
&\left.\qquad\qquad\qquad\qquad\qquad\quad-e^{-(A+1)\tilde{\Gamma}_{(k+1)}}\sum\limits_{j=0}^{A-1}{\frac{\left(\tilde{\Gamma}_{(k+1)}\right)^j}{j!}}
\right),
\end{split}
\end{equation}
where $\tilde{\Gamma}_{(i)}=\Gamma_{(i)}/\rho$, $i=1,\cdots,L$.
\par
The $A=N$ case is treated separately from the $A<N$ case because no
relays back off in this case. The probability that all relays and
the $k$-th MS are selected can be derived as
\begin{equation}
\mathrm{Pr}\left(N,MS_{(k)}\right)
=e^{-(N+1)\tilde{\Gamma}_{(k)}}\sum\limits_{j=0}^{N}{\frac{\tilde{\Gamma}_{(k)}^j}{j!}}
-e^{-(N+1)\tilde{\Gamma}_{(k+1)}}\sum\limits_{j=0}^{N}{\frac{\tilde{\Gamma}_{(k+1)}^j}{j!}}.
\end{equation}
The bound of the SER of the proposed scheme is given by,
\begin{equation}
\label{ser} 0<P_e\left(\mathcal{A},MS_{(k)}\right)\leq SER_{tgt}.
\end{equation}
Substituting (\ref{pr_a})--(\ref{ser}) into (\ref{throughput_eq})
yields the upper bound and lower bound of the throughput for the
proposed scheme,
\begin{equation}
\label{throughput_neq} \zeta_{upper}=
\sum\limits_{k=1}^{L}{\sum\limits_{A=1}^{N} {k \cdot
\mathrm{Pr}\left(A,MS_{(k)}\right)}},
\end{equation}
and
\begin{equation}
\label{lower_bound} \zeta _{lower}=  \sum\limits_{k=1}^{L}{\left(k
\cdot(1- SER_{tgt})^{L_{bit}/2k} \sum\limits_{A=1}^{N} {
\mathrm{Pr}\left(A,MS_{(k)}\right)}\right)}.
\end{equation}

\subsection{Numerical Results}
Numerical results are provided in this subsection. In all results,
the horizontal axis indicates the SNR of the source-relay and
relay-destination channels without fading, and $L_{bit}=1000$. The
candidate modulation schemes used in the simulation are shown in
TABLE \ref{table_1}, while the SER target is $SER_{tgt}=10^{-4}$.
\begin{table} [h]
\centering \caption{\label{table_1}Candidate MSs and Corresponding SNR Threshold }
\begin{tabular}[t]{c|c}
\hline MS & Threshold\\
\hline
No transmission & -\\
\hline
QPSK & 11.80 dB \\
\hline
16QAM & 19.00 dB\\
\hline
64QAM & 25.32 dB\\
\hline
256QAM & 31.42dB\\
\hline
\end{tabular}
\end{table}

\begin{figure}
\centering
\includegraphics[width=3in]{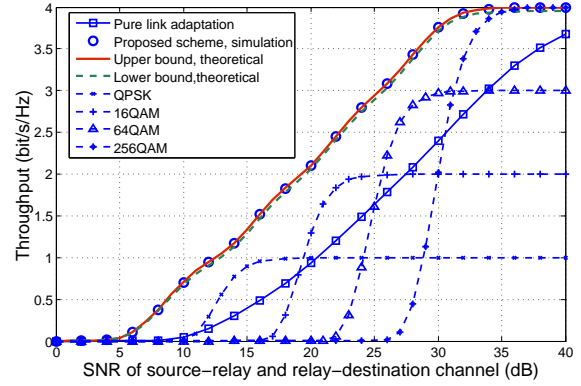}
\caption{ Throughput performance of 5-relay networks. Six cases are
studied: the proposed scheme, link adaptation with no relay
selection, QPSK with relay selection, 16QAM with relay selection,
64QAM with relay selection and 256QAM with relay selection.}
\label{throughput}
\end{figure}

In Fig. \ref{throughput}, the throughput in a network with 5
candidate relays is presented. It is observed that the proposed
scheme increases the throughput relative to the link adaptation
scheme without relay selection (with all candidate relays
participating in the second step transmission) by over $0.5$
bit/s/Hz in the SNR range 10-35 dB. This is reasonable because relay
selection can mitigate the risks that the channel is in deep fading
by utilizing diversity. Also, the proposed scheme outperforms pure
relay selection scheme with fixed MS. This is what should be
expected from the proposed scheme because with the same average SNR
at the destination, the proposed scheme can adaptively switch to a
low or high spectral efficiency MS according to the channel
realizations, so that the throughput is always maximized. Fig.
\ref{throughput} also indicates that the bounds derived in
(\ref{throughput_neq}) and (\ref{lower_bound}) are very close to the
simulation results.

\begin{figure}
\centering
\includegraphics[width=3in]{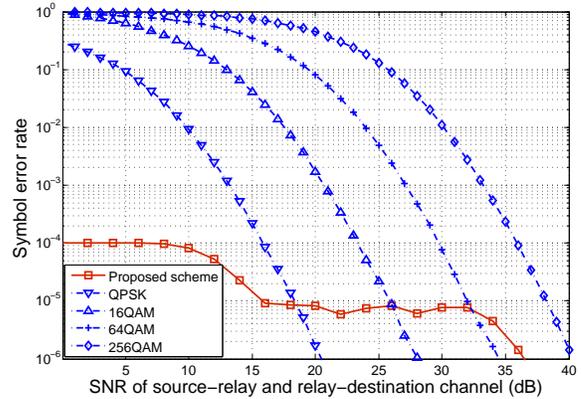}
\caption{ Average SER of 5-relay networks. Five cases are studied:
the proposed scheme, QPSK with relay selection, 16QAM with relay
selection, 64QAM with relay selection and 256QAM with relay
selection.} \label{ber}
\end{figure}

\par
Fig. \ref{ber} shows the average SER of the proposed scheme in a
5-relay network. As is expected, the SER of the proposed scheme is
always below the $SER_{tgt}$. While a very low SER value may be not
desirable from the point of the system level, maintaining it at an
appropriate level, if not too high, is beneficial to maximize the
system throughput. This is clear if we re-examine the spectral
efficiency performance in Fig. \ref{throughput}. Although the SER of
the proposed scheme is larger than relay selection with lower
modulation rank in high SNRs, the throughput performance of the
proposed scheme is improved.

\subsection{Signaling Comparison of the General Scheme (Section III-A) and Improved Scheme (Section III-B)}
In the general scheme, feedback signaling channel was pre-scheduled by scheduler in higher layer.
Prior to scheduling, none of the nodes know what relays will be actually selected, which is based
on instantaneous channel and packet detection error. So the signaling channel must be scheduled for
all possible relays, both selected relays and not-selected relays. Since the feedbacks from destination
to relays are in each relay's dedicated channel, the number of signaling bits is proportional to the number of relays.
\par
However, in the improved scheme, signaling bits which indicate the link adaptation and relay selection results are
transmitted in a broadcast channel. Here we calculate the number of signaling bits of the general scheme and proposed
scheme for comparison, as shown in Table 2.
\par
\begin{table} [ht]
\centering \caption{\label{table_2}Number of Signaling bits}
\begin{tabularx}{250pt}{|>{\hsize=.35\hsize}X|>{\hsize=1.3\hsize}X|>{\hsize=1.35\hsize}X|}
\hline
& TDD system & FDD system\\
\hline
General scheme & $N\times$(MS+1\_bit+Eg)+MS $\quad\quad =9N + (N+1)\lceil\log_2(L)\rceil$ &
$N\times$(MS+1\_bit+BWI)+MS $\quad=17N + (N+1)\lceil\log_2(L)\rceil$
\\
\hline
Proposed scheme& MS+Eg $=8+\lceil\log_2(L)\rceil$ & MS+$N\times$BWI= $16N+\lceil\log_2(L)\rceil$\\
\hline
\end{tabularx}
\end{table}

In Table 2, $N$ is the number of relays, including both selected
relays and not-selected relays. MS is the index of the selected MS,
which is $\lceil\log_2(L)\rceil$ bits. The 1\_bit in general scheme
is the signaling bit that indicates each specific relay whether it
is selected or not, and is assumed 1 bit. Eg is the summation of
gaining factors. It is a real number and is assumed 8 bits. The BWI
bits are used for both the two schemes in FDD system to indicate
each relay the beamforming weight. They are assumed 16 bits, as they
include the magnitude and phase. In the general scheme, the
destination also needs to feedback MS to the source in the dedicated
channel. By using the MS selection results to indicate the relay
selection results, the bits used in general scheme for relay
selection results feedback are saved, and the signaling overhead of
relay selection and link adaptation is significantly reduced.

\section{Conclusion}
\label{conclusion} Motivated by the improvement in throughput and
spectral efficiency afforded by link adaptation and relay selection,
an approach for joint relay selection and link adaptation is
proposed in this paper. Since the relay selection tends to select
the links with higher capacity while the link adaptation makes
efficient use of those links, the combination of relay selection and
link adaptation can further provides synergistic gains. To reduce
the signaling overhead of relay selection and link adaptation, a
novel signaling mechanism is proposed, along with a simple and
optimal selection algorithm. The bounds of the received SNR and the
throughput is derived. Simulation results further confirms that the
proposed scheme can improve the throughput and SER performance
compared with pure link adaptation and relay selection schemes.

\appendices
\section{Closed-Form Expression for $f(u,v)$}
\label{appendixA}
Define $Z$ as $Z=\sum\limits_{i=1}^{A}\gamma_{2,i}$, and the pdf of $Z$ is denoted by $g(z)$. Since $\gamma_{2,i}$ are independent exponential random variables, the
Laplace transform, $\mathcal {L}g$, of $g_{Z}(z)$ is given by $\mathcal {L}g \left(s\right)=\frac{1}{\left(1+s\right)^A}$. Hence the pdf of $Z$ can be expressed as
\begin{equation}
\label{app1}
 g(z)=\frac{z^{(A-1)}e^{-\frac{z}{\rho}}}{\rho^A(A-1)!},
\end{equation}
and the cumulative distribution function (cdf) of $Z$ is given by
\begin{equation}
\label{app2} G_{Z}(z)=1-\sum\nolimits_{i=0}^{A-1}\frac{z^i
e^{-\frac{z}{\rho}}}{\rho^i i!}.
\end{equation}
The joint pdf of $\gamma_{[A]}$ and $\gamma_{[A+1]}$ can be derived
using the theory of ``order statistics'' \cite{order_sta} as
\begin{equation}
\psi_{\gamma_{[A]},\gamma_{[A+1]}}(x,y)=
\frac{N!e^{-\frac{y}{\rho}}e^{-\frac{Ax}{\rho}}\left(1-e^{-\frac{y}{\rho}}\right)^{N-A-1}}{\rho^2 (N-A-1)!(A-1)!} ,\quad0<y<x.
\end{equation}
By ``order statistics'', we also have
\begin{equation}
\label{app3}
\psi_{\gamma_{[A+1]}}(y) = \frac{N!e^{-\frac{(A+1)y}{\rho}}\left(1-e^{-\frac{y}{\rho}}\right)^{N-A-1}}{\rho(N-A-1)!A!},\quad\quad y>0,
\end{equation}
where $\psi_{\gamma_{[A+1]}}(y)$ is the pdf of $\gamma_{[A+1]}$. Thus, the joint cdf of $U_A$ and $V_A$ can be expressed as
\begin{equation}
\begin{split}
&F(u,v)
=
\begin{cases}
\mathrm{Pr}\left(Z>u,\gamma_{[A]}<u,\gamma_{[A+1]}<v\right)\\
\qquad\quad +\mathrm{Pr}\left(Z<u,\gamma_{[A+1]}<v\right), &  0<v<u\\
\mathrm{Pr}\left(Z<u,\gamma_{[A+1]}<v\right), &  0<u<v
\end{cases}\\
&=
\begin{cases}
\left(1-G(u)\right)\Psi(u,v)+G(u)\Psi_{\gamma_{[A+1]}}(v),& \quad0<v<u\\
G(u)\Psi_{\gamma_{[A+1]}}(v), &\quad0<u<v,
\end{cases}
\end{split}
\end{equation}
where $\Psi(u,v)$ denotes the joint cdf of $\gamma_{[A]}$ and
$\gamma_{[A+1]}$. Thus the joint pdf of $U_A$ and $V_A$ can be derived
as
\begin{equation}
\label{app4}
f(u,v)=
\begin{cases}
g(u)\psi_{\gamma_{[A+1]}}(v)- G(u)\psi (u,v) \\
\qquad\qquad-g(u) \frac{{\partial \Psi (u,v)}}{{\partial v}}+\psi(u,v), &0<v<u\\
g(u)\psi_{\gamma_{[A+1]}}(v) ,&0<u<v,
\end{cases}
\end{equation}
where
\begin{equation}
\label{app5}
\frac{{\partial \Psi (u,v)}}{{\partial v}} = \frac{{N!e^{ - \frac{v}{\rho}}
\left(1 - e^{ - \frac{v}{\rho}} \right)^{N - A - 1} \left(1 - e^{ - \frac{Au}{\rho}} \right)
}}{{\rho(N - A - 1)!A!}}.
\end{equation}
The closed-form expression for $f(u,v)$ can be derived by
substituting (\ref{app1})-(\ref{app3}) and (\ref{app5})into (\ref{app4}).






\begin{thebibliography}{20}
\bibitem{Laneman}
J.~N.~Laneman, David.~N.~C.~Tse, and G.~W.~Wornell, ``Cooperative
Diversity in Wireless Networks: Efficient Protocols and Outage
Behavior,'' \emph{IEEE Trans. Inform. Theory}, vol. 50, no.12, pp.
3062--3080, Dec. 2004.
\bibitem{patent}
R.~Madan, N.~B.~Mehta, A.~F.~Molisch, and J.~Zhang,
``Energy-Efficient Cooperative Relaying Over Fading Channels With
Simple Relay Selection,''
 \emph{IEEE Trans. Wireless Commun.}, vol. 7, no. 8, pp. 3013--3025, Aug. 2008.


\bibitem{GoldSmith_AMC}
A.~Goldsmith and S.~G.~Chua, ``Adaptive coded modulation for fading
channels,'' \emph{IEEE Trans. Commun.}, vol. 46, no. 5, pp.
595--602, May 1998.

\bibitem{Zinan_GC}
Z.~Lin, E.~Erkip, and M.~Ghosh, ``Rate adaptation for cooperative
systems,'' \emph{IEEE GlobeCom}, San Francisco, USA, Nov. 2006, pp.
1--5.



\bibitem{single_multiple_relay_selection}
Y.~Jing, and H.~Jafarkhani, ``Single and multiple relay selection schemes
 and their achievable diversity orders,'' \emph{IEEE Trans. Wireless Commun.},
 vol. 8, no. 3, pp. 1414--1423,  Mar. 2009.
\bibitem{A_Bletas}
A.~Bletsas, A.~Khisti, D.~P.~Reed, and A.~Lippman, ``A simple
cooperative diversity method based on network path selection,''
\emph{IEEE J. Select. Areas Commun.}, vol. 24, no. 3, pp. 659--672,
Mar. 2006.


\bibitem{Jing_AF}
Y.~Jing and H.~ Jafakhani, ``Network beamforming using relays with
perfect channel information,'' \emph{IEEE Trans. Inf. Theory}, vol.
55, no. 6, pp. 2499--2517, Jun. 2009.

\bibitem{distributed_beamforming_Comm_Mag}
R.~Mudumbai, D.~R.~Brown~III, U.~Madhow, and H.~V.~Poor,
``Distributed transmit
 beamforming: challenges and recent progress,'' \emph{IEEE Commun. Mag.}, vol. 47, no. 2,
 pp. 102--110, Feb. 2009.

 \bibitem{Jing_YD_bit}
E.~Koyuncu, Y.~Jing, and H.~Jafarkhani, ``Distributed beamforming in
wireless relay networks with quantized feedback,'' \emph{IEEE J.
Select. Areas Commun.}, vol. 26, no. 8, pp. 1429--1439, Oct. 2008.

\bibitem{dis_BF_DF}
L.~Wang, C.~Zhang, J.~Zhang, and G.~Wei, ``Distributed beamforming
with limited feedback in regenerative cooperative networks,'' in
\emph{IEEE WiCom}, Beijing, China, Sep. 2009, pp. 1--4.

\bibitem{Jonit_Rate}
M.~R.~Souryal and N.~Moayeri, ``Joint Rate Adaptation and
Channel-Adaptive Relaying in 802.11 Ad Hoc Networks,'' in \emph{IEEE
Military Communications Conference}, Washington D.C., USA, Oct.
2006, pp. 1--8.



\bibitem{order_sta}
P.~J.~Bickel and K.~Doksum, \emph{Mathematical Statistics: Basic
Ideas and Selected Topics}, first ed. Oakland, CA: Holden-Day, 1977.




\end{thebibliography}
%

\end{document}